\documentclass[sigconf]{acmart}
\usepackage{hyperref}

\AtBeginDocument{%
  }

\setcopyright{acmlicensed}
\copyrightyear{2018}
\acmYear{2018}
\acmDOI{XXXXXXX.XXXXXXX}
\acmConference[Conference acronym 'XX]{Make sure to enter the correct
  conference title from your rights confirmation email}{June 03--05,
  2018}{Woodstock, NY}
\acmISBN{978-1-4503-XXXX-X/2018/06}

\begin{document}

\title{Quadratic Interest Network for Multimodal Click-Through Rate Prediction}

\author{Honghao Li}
\email{salmon1802li@gmail.com}
\orcid{0009-0000-6818-7834}
\affiliation{%
  \institution{Anhui University}
  \city{Hefei}
  \state{Anhui Province}
  \country{China}
}

\author{Hanwei Li}
\email{lihanwei@stu.ahu.edu.cn}
\affiliation{%
  \institution{Anhui University}
  \city{Hefei}
  \state{Anhui Province}
  \country{China}
}

\author{Jing Zhang}
\email{e23301310@stu.ahu.edu.cn}
\affiliation{%
  \institution{Anhui University}
  \city{Hefei}
  \state{Anhui Province}
  \country{China}
}

\author{Yi Zhang}
\email{zhangyi.ahu@gmail.com}
\affiliation{%
  \institution{Anhui University}
  \city{Hefei}
  \state{Anhui Province}
  \country{China}
}

\author{Ziniu Yu}
\email{sanglei@ahu.edu.cn}
\affiliation{%
  \institution{Anhui University}
  \city{Hefei}
  \state{Anhui Province}
  \country{China}
}

\author{Lei Sang}
\email{sanglei@ahu.edu.cn}
\affiliation{%
  \institution{Anhui University}
  \city{Hefei}
  \state{Anhui Province}
  \country{China}
}

\author{Yiwen Zhang}
\authornote{Corresponding author}
\email{zhangyiwen@ahu.edu.cn}
\affiliation{%
  \institution{Anhui University}
  \city{Hefei}
  \state{Anhui Province}
  \country{China}
}

\renewcommand{\shortauthors}{Li et al.}

\begin{abstract}
  Multimodal click-through rate (CTR) prediction is a key technique in industrial recommender systems. It leverages heterogeneous modalities such as text, images, and behavioral logs to capture high-order feature interactions between users and items, thereby enhancing the system's understanding of user interests and its ability to predict click behavior. The primary challenge in this field lies in effectively utilizing the rich semantic information from multiple modalities while satisfying the low-latency requirements of online inference in real-world applications. To foster progress in this area, the Multimodal CTR Prediction Challenge Track of the WWW 2025 EReL@MIR Workshop \cite{WWW2025challenge} formulates the problem into two tasks: (1) Task 1 of Multimodal Item Embedding: this task aims to explore multimodal information extraction and item representation learning methods that enhance recommendation tasks; and (2) Task 2 of Multimodal CTR Prediction: this task aims to explore what multimodal recommendation model can effectively leverage multimodal embedding features and achieve better performance. 

  In this paper, we propose a novel model for Task 2, named Quadratic Interest Network (QIN) for Multimodal CTR Prediction. Specifically, QIN employs adaptive sparse target attention to extract multimodal user behavior features, and leverages Quadratic Neural Networks to capture high-order feature interactions. As a result, QIN achieved an AUC of 0.9798 on the leaderboard and ranked second in the competition. The model code, training logs, hyperparameter configurations, and checkpoints are available at \url{https://github.com/salmon1802/QIN}.
\end{abstract}



\received{20 February 2007}
\received[revised]{12 March 2009}
\received[accepted]{5 June 2009}

\maketitle

\section{Introduction}

\begin{table}[h]
\caption{Top-5 teams on the leaderboard}
\begin{tabular}{cc}
\hline
Rank          & Team Name \\ \hline
1st Place     & momo      \\
2nd Place     & DISCO.AHU \\
3rd Place     & jzzx.NTU  \\
Task 1 Winner & delorean  \\
Task 2 Winner & zhou123   \\ \hline
\end{tabular}
\label{leaderboard}
\end{table}

Recommender systems have become indispensable tools for helping users navigate massive volumes of content and products. At the core of such systems lies click-through rate (CTR) prediction, which aims to estimate the probability that a user clicks on a recommended item. Accurate CTR prediction not only enhances user experience by delivering more relevant recommendations, but also drives commercial success by increasing engagement and conversion rates on online platforms such as e-commerce, social media, and content streaming services.

Traditional CTR prediction models primarily rely on log-based data \cite{FINAL, openbenchmark, CETN, FCN, adagin}, including user profiles, item attributes, and contextual features, to model user behavior. With the growing availability of multimodal data—such as textual descriptions, images, and videos—it becomes increasingly feasible to leverage richer semantic information to improve prediction accuracy. Multimodal CTR prediction \cite{SimCEN, liu2023prior} seeks to integrate these diverse sources to better understand user interests and item features, thereby improving recommendation quality. However, a key challenge remains: how to incorporate rich multimodal information into CTR models while satisfying the low-latency requirements of online inference in industrial applications.

\begin{figure*}[t]
    \centering
    \begin{minipage}[t]{0.9\linewidth}
    \includegraphics[width=\linewidth]{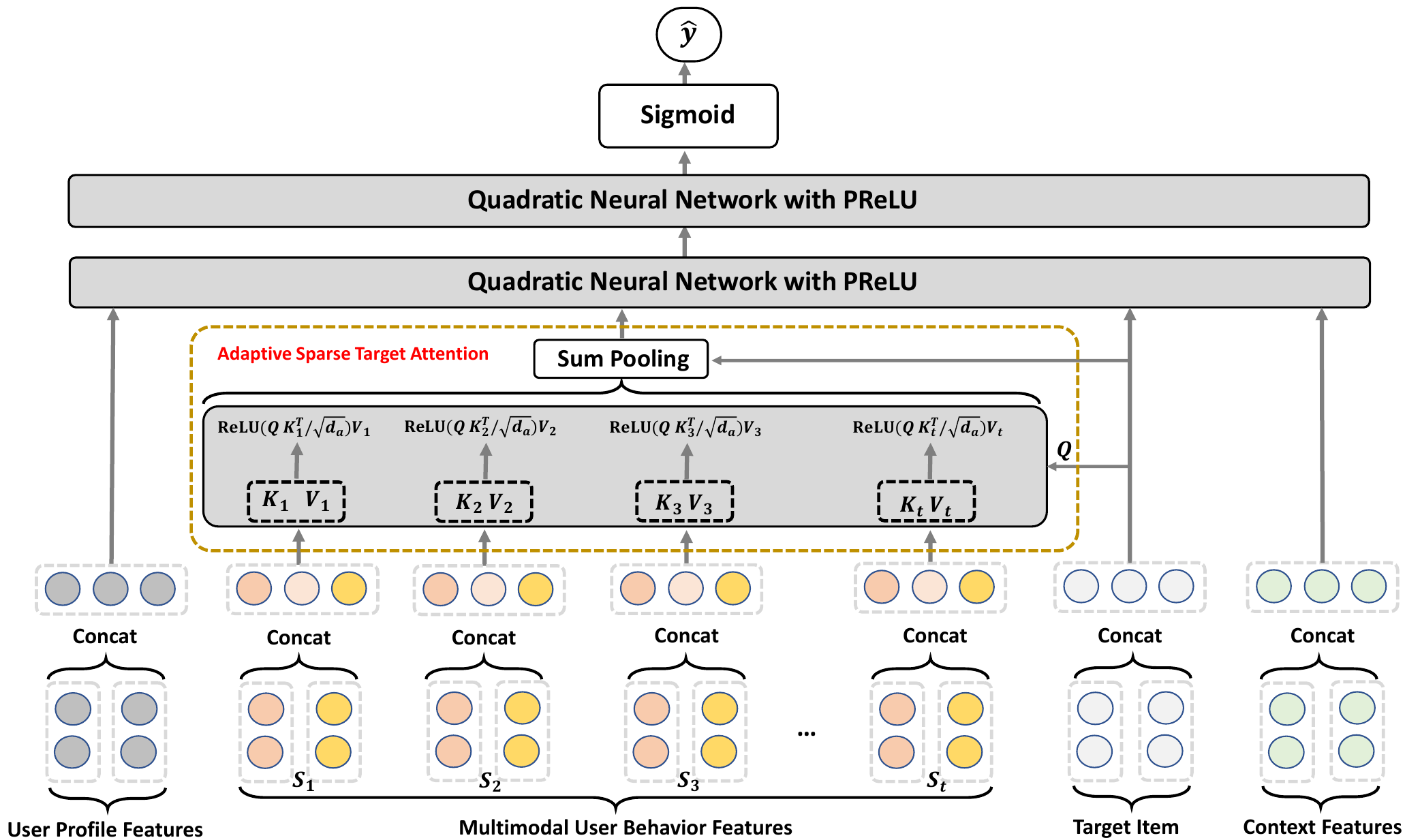}
    \end{minipage}
    \captionsetup{justification=raggedright, font=small}
    \caption{The architecture of Quadratic Interest Network.}
    \label{QIN}
\end{figure*}

To foster progress in this emerging field, the Multimodal CTR Prediction Challenge Track of the WWW 2025 EReL@MIR Workshop \cite{WWW2025challenge} is introduced. The challenge defines two tasks: (1) Multimodal Item Embedding, which explores how to extract and represent multimodal item features to enhance recommendation performance; and (2) Multimodal CTR Prediction, which investigates how to effectively utilize multimodal embeddings in CTR models to achieve superior performance.

In this paper, we focus on Task 2 of the challenge. Our goal is to design a multimodal recommendation model that effectively leverages multimodal embeddings to enhance CTR prediction. To this end, we propose a novel deep learning architecture: the Quadratic Interest Network (QIN) for Multimodal Click-Through Rate Prediction. QIN incorporates two key innovations to address the challenges of multimodal CTR prediction: (1) it adopts an adaptive sparse target attention (ASTA) mechanism to extract multimodal user behavior features. This mechanism enables the model to dynamically focus on the most informative parts of users' interaction histories, which is critical for handling the complexity and diversity of multimodal data; (2) QIN employs Quadratic Neural Networks (QNN) to capture high-order feature interactions. In CTR prediction, user click behavior often depends on the interplay of multiple features, such as the combination of past purchase history and the visual appeal of an item. By explicitly modeling these interactions, QIN learns more expressive representations that better reflect the underlying factors influencing user decisions. Our experimental results demonstrate the effectiveness of QIN. As shown in Table \ref{leaderboard}\footnote{\url{https://erel-mir.github.io/challenge/results/}}, on the challenge dataset\footnote{\url{https://recsys.westlake.edu.cn/MicroLens_1M_MMCTR}}, QIN achieves an AUC of 0.9798, ranking second on the competition leaderboard. This strong performance highlights the practical potential of QIN in industrial recommender systems.

\section{Quadratic Interest Network}
As illustrated in Figure \ref{QIN}, the proposed Quadratic Interest Network (QIN) primarily consists of two components: Adaptive Sparse Target Attention and a Quadratic Neural Network. In this section, we introduce the model in a bottom-up manner.

\subsection{Adaptive Sparse Target Attention}
Target attention is a widely adopted approach for modeling user behavior and extracting user interests~\cite{DIN, SDIM, TWIN, MIRRN}. It treats the target item as the query and the user behavior sequence as keys and values, applying an attention mechanism to dynamically assign weights to each item in the sequence. This allows the model to focus on behaviors that are most relevant to the target item. Specifically, let the target item embedding be denoted as $\mathbf{x}_t \in \mathbb{R}^{d_t}$, and the user behavior sequence embeddings as $\mathbf{x}_b \in \mathbb{R}^{S \times d_b}$, where $d_t$ and $d_b$ represent the embedding dimensions of the target item and behavior features, respectively, and $S$ denotes the length of the behavior sequence. The formulation of target attention is defined as follows:
\begin{equation}  
\begin{aligned}
\mathbf{Q} = \mathbf{W}_q\mathbf{x}_t , \quad \mathbf{K} = \mathbf{W}_k\mathbf{x}_b , \quad \mathbf{V} = \mathbf{W}_v\mathbf{x}_b ,
\end{aligned}
\end{equation}
where $\mathbf{W}_q \in \mathbb{R}^{d_a \times d_t}$, $\mathbf{W}_k \in \mathbb{R}^{d_a \times d_b}$, $d_a$ denotes the attention dimension, and $\mathbf{W}_v \in \mathbb{R}^{d_a \times d_b}$ are learnable projection matrices that transform the target item and behavior sequence embeddings into the query ($\mathbf{Q} \in \mathbb{R}^{1 \times d_a}$), key ($\mathbf{K} \in \mathbb{R}^{S \times d_a}$), and value ($\mathbf{V} \in \mathbb{R}^{S \times d_a}$), respectively. The attention weights are computed as:
\begin{equation}  
\begin{aligned}
\mathbf{o} = \text{SoftMax}\left( \frac{\mathbf{Q} \mathbf{K}^\top}{\sqrt{d_a}} \right)\mathbf{V},
\end{aligned}
\end{equation}
where $\sqrt{d_k}$ is a scaling factor to stabilize the dot-product attention, $\mathbf{o} \in \mathbb{R}^{d_v}$ is the aggregated user interest representation, capturing the weighted combination of behavior sequence features most relevant to the target item. This QKV-based target attention mechanism enables the CTR model to effectively capture user preferences by focusing on the most pertinent historical interactions, thus enhancing performance.

However, the normalized soft attention mechanism using SoftMax is not suitable for streaming recommendation scenarios \cite{DIN, HSTU}, particularly in click-through rate (CTR) prediction tasks with high real-time requirements, long user behavior sequences, and multimodal features. The SoftMax operation enforces global normalization of attention weights across the entire behavior sequence, which not only increases computational complexity but also risks diluting focus on critical behaviors, especially when the sequence contains significant noise or irrelevant actions. To address this issue, we propose Adaptive Sparse Target Attention (ASTA), which replaces SoftMax with ReLU to generate non-normalized hard attention in a simple yet effective manner, thereby improving prediction accuracy while reducing computational overhead.
\begin{equation}  
\begin{aligned}
\mathbf{o} = \text{ReLU}\left( \frac{\mathbf{Q} \mathbf{K}^\top}{\sqrt{d_a}} \right)\mathbf{V} + \mathbf{x}_t,
\end{aligned}
\end{equation}
Notably, we find that computing attention weights with ASTA eliminates the need for additional Dropout \cite{dropout} to prevent overfitting in the attention network, as confirmed by our ablation studies. This is likely because ReLU-based attention weights are sufficiently significant and sparse, rendering additional regularization methods detrimental to the model's ability to capture critical user behavior patterns. Additionally, we explore ReLU$^2$ \cite{relu2} and SiLU \cite{HSTU} as alternatives, but their performance is inferior to the simple yet effective ReLU.

\subsection{Quadratic Neural Network}
\begin{figure}[t]
    \centering
    \begin{minipage}[t]{0.8\linewidth}
    \includegraphics[width=\linewidth]{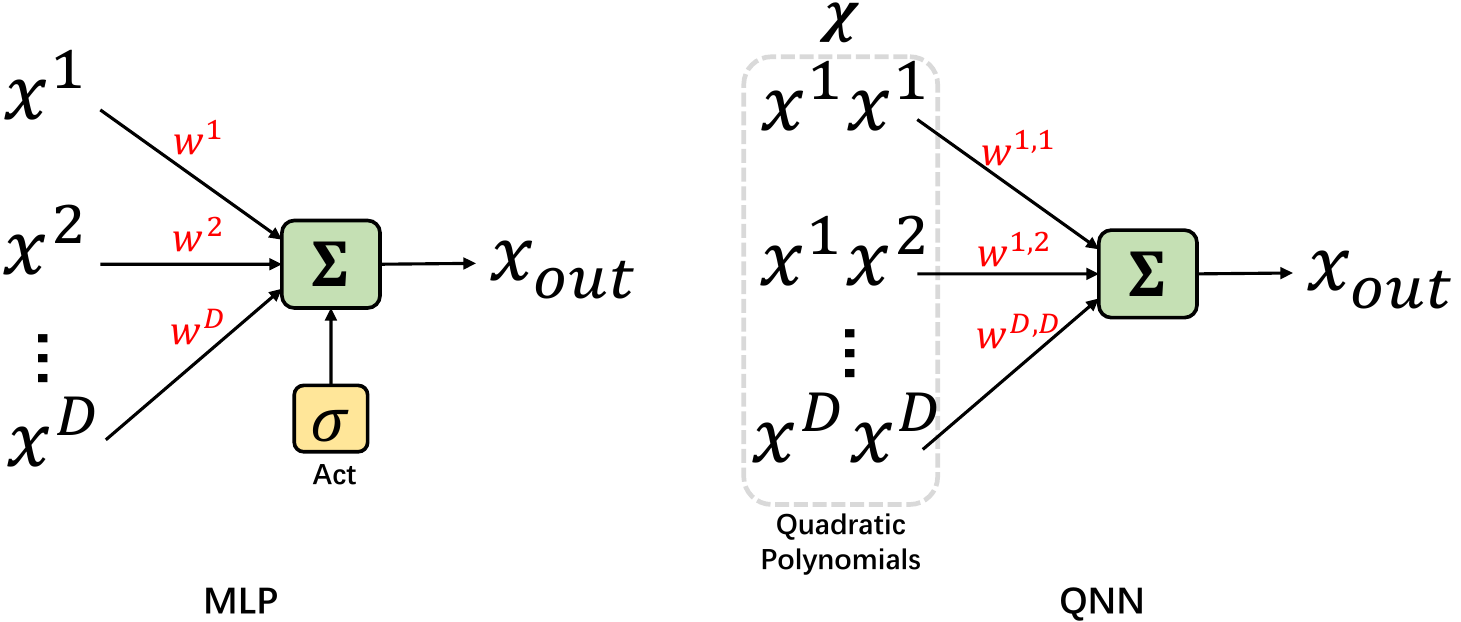}
    \end{minipage}
    \captionsetup{justification=raggedright, font=small}
    \caption{Comparison of MLP and QNN. The input to the MLP consists of raw features, while the QNN uses linearly independent quadratic polynomials as input.}
    \label{MLPvsQNN}
    \vspace{-1em}
\end{figure}

Quadratic Neural Networks leverage quadratic interaction terms to model complex feature relationships, serving as a powerful alternative to widely used architectures such as CrossNet \cite{dcn, dcnv2, FCN} and MLP \cite{finalmlp, DNN}. As illustrated in Figure \ref{MLPvsQNN}, MLPs take raw features $x_i \in \mathbf{x}$ as input, modeling nonlinear feature relationships through linear weighting and activation functions. In contrast, QNNs directly utilize linearly independent quadratic polynomials as input, enabling explicit capture of high-order feature interactions. This approach enhances QNNs' expressive power in handling complex feature relationships, leading to more accurate predictions of user click behavior.

Specifically, in QNN, we adopt an implementation based on the Khatri–Rao product \cite{Khatri-Rao} combined with a post-activation function (in some cases, the mid-activation function performs better). Let $\mathbf{X}_1 \in \mathbb{R}^{D}$ denote the first-order feature interaction vector obtained by concatenating all features. The recursive formulation of QNN is given by (the residual term is omitted):
\begin{equation}
\label{T26}
\begin{aligned}
\mathbf{X}_{l+1} &= \sigma\left(\mathbf{X}_l \bullet \mathbf{W}_l \mathbf{X}_l\right),\ \ l=1,2,\dots,L, \\
\sigma\left(x_{l+1}^i\right) &\overset{l\ >\ 1}{=} \sigma\left(x_{l-1}^i \sum_{j=1}^{D}\sum_{p=1}^{M} w_{l-1}^{p,i,j} x_{l-1}^j \sum_{k=1}^{D}\sum_{q=1}^{M} w_l^{q,i,k} x_l^k\right), \\
&= \underbrace{\sigma\left(A_{(1,1)} x_{l-1}^1 x_l^1 + A_{(1,2)} x_{l-1}^1 x_l^2 + \cdots + A_{(D,D)} x_{l-1}^D x_l^D\right)}_{D^2 \text { interaction items}},
\end{aligned}
\end{equation}
where $x_{l}^j \in \mathbf{X}_l$ denotes the $j$-th element of $\mathbf{X}_l$, $\sigma$ denotes the post-activation function, in this paper the PReLU \cite{PReLU} commonly used in sequence modeling is chosen \cite{DIEN, MIRRN}, $\bullet$ denotes Khatri–Rao product, $\mathbf{W}_l \in \mathbb{R}^{M \times D \times D}$, $A_{(j, k)} = \sum_{p=1}^{M}\sum_{q=1}^{M} w_{l-1}^{p,i,j} w_{l}^{q,i,k} x_{l-1}^i$, and $M$ as a hyperparameter to adjust the model capacity and enhance scalability. In this manner, QNN captures high-order feature interactions layer by layer while retaining some of its inherent capability for smooth nonlinear approximation.

For further details on QNN, refer to our forthcoming paper, which is currently under preparation. Upon completion and acceptance, we will provide the paper's link \href{https://github.com/salmon1802/QNN}{\color{red}{here}}. 

\subsection{Training}
We employ a simple linear layer to transform the QNN output into a logit, followed by a Sigmoid function to obtain a binary classification probability prediction. For loss computation, we adopt the widely used binary cross-entropy loss function \cite{openbenchmark, TF4CTR} to effectively measure the discrepancy between the model's predicted probabilities and the true labels.
\begin{equation}
\begin{aligned}
\hat{y} &= \text{Sigmoid}\left(\mathbf{W}\mathbf{X}_L + \mathbf{b}\right), \\
\mathcal{L}_{\scriptscriptstyle CTR} &=-\frac{1}{N} \sum_{i=1}^N\left(y_i \log \left(\hat{y}_i\right)+\left(1-y_i\right) \log \left(1-\hat{y}_i\right)\right)
\end{aligned}
\end{equation}
where $N$ denotes the batch size, and $\mathbf{X}_L$ denotes the output of the last layer of QNN. Notably, during this challenge, we experiment with various loss augmentation methods, including BCE + BPR Loss \cite{rankandlog}, Focal Loss \cite{focalloss}, and Contrastive Loss \cite{simGCL}. However, these methods yield marginal improvements.

\section{Experiments}
\subsection{Implementation Details}
All experiments are conducted on an NVIDIA GeForce RTX 4090 GPU, with our implementation based on the FuxiCTR framework \cite{FuxiCTR}. We use a learning rate of 2e-3, an embedding weight decay strength of 2e-4, a batch size of 8192, and a feature embedding dimension of 128 for all features. For QNN, we set $L = M = 4$ and a Dropout rate of 0.1. Further details on model hyperparameters and dataset configurations are available in our straightforward and accessible running logs\footnote{\url{https://github.com/salmon1802/QIN}}.

\subsection{Performance Comparison}
DIN \cite{DIN} is a classic CTR prediction model based on user behavior sequences and serves as the official baseline for this challenge. Consequently, we compare QIN against DIN. As shown in Table \ref{baseline}, QIN achieves an AUC score improvement of 0.1046 over DIN and ranks second on the competition leaderboard. These results demonstrate the effectiveness of QIN.

\begin{table}[t]
\caption{Performance Comparison with DIN}
\begin{tabular}{cc}
\hline
Methods  & AUC on valid set \\ \hline
DIN  & 0.8655 \\
QIN     & 0.9701      \\ \hline
\end{tabular}
\label{baseline}
\end{table}

\subsection{Ablation Study}

\begin{table}[t]
\caption{Ablation Study of QIN.}
\begin{tabular}{cc}
\hline
Variety  & AUC on valid set \\ \hline
QIN w/o QNN  & 0.7396 \\
QIN w/o ASTA & 0.9321 \\
ASTA w/ SoftMax & 0.9490  \\
QNN w/o PReLU & 0.9584  \\ 
ASTA w/ Dropout & 0.9681 \\ \hline
QIN     & 0.9701      \\ \hline
\end{tabular}
\label{ablation}
\end{table}

To evaluate the effectiveness of individual components in QIN, we conduct ablation studies:
\begin{itemize}
\item QIN w/o QNN: QIN employs an MLP with structure [1024, 512, 256] instead of QNN.
\item QIN w/o ASTA: QIN uses mean pooling instead of ASTA.
\item ASTA w/ SoftMax: ASTA adopts SoftMax instead of ReLU.
\item QNN w/o PReLU: QNN uses ReLU instead of PReLU.
\item ASTA w/ Dropout: ASTA incorporates Dropout with a rate of 0.1.
\end{itemize}

The results, presented in Table \ref{ablation}, demonstrate that QNN is a highly effective method for modeling feature interactions, significantly enhancing QIN's performance. Specifically, removing the QNN component (QIN w/o QNN) leads to a substantial AUC drop on the validation set from 0.9701 (full QIN) to 0.7396, underscoring QNN's critical role in capturing complex feature relationships. Additionally, removing the Adaptive Sparse Target Attention (ASTA) component (QIN w/o ASTA) reduces the AUC to 0.9321, highlighting ASTA's effectiveness in extracting user interests. Further analysis reveals that replacing ASTA's sparse attention with SoftMax (ASTA w/ SoftMax) lowers the AUC to 0.9490, confirming the advantage of sparse attention in reducing noise and improving model generalization. Moreover, the use of PReLU activation in QNN contributes to performance, as its removal (QNN w/o PReLU) results in an AUC of 0.9584. Finally, adding Dropout to ASTA (ASTA w/ Dropout) yields an AUC of 0.9681, slightly below the full QIN's 0.9701, indicating that ASTA's sparse nature obviates the need for additional Dropout regularization. In summary, QIN achieves superior CTR prediction performance through the synergistic interplay of QNN and ASTA.

\section{Conclusion}
In this paper, we introduced a novel Quadratic Interest Network for Multimodal Click-Through Rate Prediction, which primarily consisted of an Adaptive Sparse Target Attention module and a Quadratic Neural Network. The model effectively leveraged multimodal features to model user interests and capture feature interactions, thereby improving performance. It achieved second place in the Multimodal CTR Prediction Challenge Track of the WWW 2025 EReL@MIR Workshop, demonstrating its effectiveness and potential for industrial applications. 

We would be deeply honored if this work could provide readers with a small amount of technical insight.


% \begin{acks}
% \end{acks}

\bibliographystyle{ACM-Reference-Format}
\bibliography{sample-base}

\end{document}